# Electromagnetic polarization matrix and its physical interpretation

José J. Gil,[1*] Andreas Norrman[2], Ari T. Friberg[2], Behnaz Fazlpour[2] and Tero Setälä[2]
[1]*Independent Researcher, 29690 Casares Costa, Spain*
[2]*Center for Photonics Sciences, University of Eastern Finland, P.O. Box 111, FI-80101 Joensuu, Finland*
*Corresponding author: ppgil@unizar.es

Despite the intrinsic coupling between electric and magnetic fields in random stationary light, their polarization properties are not mutually determined. A complete second-order description thus necessitates a joint electromagnetic treatment. The 6×6 electromagnetic polarization matrix introduced here generalizes the conventional electric 3×3 matrix by incorporating both electric and magnetic contributions together with their mutual correlations. It consists of diagonal 3×3 blocks, representing the electric and magnetic polarization matrices, and off-diagonal 3×3 blocks that encode the full structure of electric–magnetic cross-correlations. The information contained in this matrix can be interpreted through physically meaningful quantities such as active and reactive energy fluxes, in-phase and quadrature alignment matrices, and global indices describing electric–magnetic coupling. The formalism is applied to a field composed of two orthogonally propagating plane waves sharing a common linear electric polarization (a simple yet physically realizable configuration) that demonstrates the need for a general combined electric–magnetic representation even in free space. This approach provides a comprehensive and unified framework for characterizing electromagnetic polarization beyond the electric-field description alone, bridging classical statistical optics with quantum-like density-matrix interpretations

## 1. Introduction

The polarization of light is commonly represented through the electric field, since its interaction with matter is typically much stronger than that of the magnetic field. Moreover, the intrinsic link between electric and magnetic fields imposed by Maxwell's equations supports this approach in many polarization phenomena. Consequently, polarization is usually described using the electric 3×3 polarization matrix (or coherency matrix), whose elements are the second-order moments of the analytic signals corresponding to the components of the local electric field [1,2]. This three-dimensional (3D) formalism allows for the description of many relevant physical situations and naturally reduces to a two-dimensional (2D) representation when the electric field evolves within a fixed plane (the so-called polarization plane) [3].

However, the conventional electric-field polarization matrix does not explicitly contain information about the energy flux or its directional balance. Even in 2D states, such quantities and other derived from the crossed electric-magnetic correlations can only be inferred under special circumstances. In contrast, the complete electromagnetic polarization matrix incorporates the electric and magnetic fields jointly, including their cross-correlation blocks, which directly describe complete second-order information of the active and reactive components of the energy flow and other manifestations of electric–magnetic coupling that are absent from the electric-only formulation.

Different electromagnetic states may share the same electric polarization matrix while exhibiting different magnetic polarization properties and electric–magnetic cross-correlations. As we show below for a simple free-space configuration of two crossing beams, only the joint electric–magnetic description captures these differences.

The idea of treating the electromagnetic field as a single six-component entity dates back to Riemann and Silberstein [4], and later to Bergman and Carozzi's 6×6 "sixtor" formalism [5], which explicitly combines the deterministic electric and magnetic fields. A related six-component representation, emphasizing "electric–magnetic democracy" in the decomposition of the Poynting vector into orbital and spin parts, was introduced by Berry in terms of a six-component state vector that stacks the suitably scaled electric and magnetic field components [6]. Six-dimensional correlation matrices have also been mentioned in other contexts, e.g., in generalized polarization-matrix treatments of optical chirality and in geometric representations of nonparaxial polarization [7,8]. Also, bispinor "wavefunction" approaches have used six-component electromagnetic state vectors to classify quadratic observables and symmetry channels, for example through the construction of an "electromagnetic symmetry sphere" for energy and momentum densities [9], or in the analysis of dual, chiral, and magnetoelectric channels in local light–matter interactions [10].

In all these formulations, second-order quantities such as energy, momentum, spin, or magnetoelectric densities are typically expressed as quadratic forms of a deterministic six-component field vector, and the corresponding 6×6 matrices are not developed as complete coherence (polarization) matrices for stationary random fields. In contrast, here we develop a comprehensive second-order statistical framework: we define the electromagnetic polarization matrix as a Hermitian positive-semidefinite 6×6 coherence matrix for random stationary light, interpret its subblocks in terms of physically measurable quantities (active/reactive energy fluxes and in-phase/quadrature alignment matrices), and introduce global correlation indices and purity–correlation relations that summarize electric–magnetic coupling in a compact and invariant way.

The aim of the present work is twofold. First, we introduce and interpret the electromagnetic polarization matrix as a compact 6×6 representation that encodes the electric and magnetic polarization properties of random stationary light in free space together with all their second-order cross-correlations, allowing each contribution to be physically identified. Second, we apply the formalism to a specific and experimentally realizable configuration, illustrating how the combined electric–magnetic treatment reveals features (such as spin, reactive flux, and directionality) that remain hidden in the electric-only framework.

## 2. Electric and magnetic polarization matrices

To establish the framework required for the introduction of the electromagnetic polarization matrix, let us first consider the 3D electric polarization matrix (also called coherency matrix) [1,2],

$$\mathbf{\Phi}_E = \langle \boldsymbol{\varepsilon}(t) \otimes \boldsymbol{\varepsilon}^{\dagger}(t) \rangle \,, \tag{1}$$

where $\boldsymbol{\varepsilon}(t) = [\varepsilon_x(t), \varepsilon_y(t), \varepsilon_z(t)]^{\mathrm{T}}$ (superscript T representing transpose matrix) is a complex electric analytic signal vector whose components are the analytic signals [11] of the electric field components [12,13] referred to a given laboratory Cartesian reference frame *XYZ*; the dagger indicates conjugate transpose, ⊗ stands for the Kronecker product, and $\langle ... \rangle$ represents time averaging. The electric polarization matrix is Hermitian and positive semidefinite, and it can be written as $\mathbf{\Phi}_E = I_E\, \hat{\mathbf{\Phi}}_E$, where $I_E = \mathrm{tr}\,\mathbf{\Phi}_E$ is the electric intensity (tr denotes the trace), and $\hat{\mathbf{\Phi}}_E$ is the dimensionless electric polarization density matrix. In SI units, the elements of $\mathbf{\Phi}_E$ (and hence $I_E$) are expressed in $\mathrm{V}^2/\mathrm{m}^2$. The quantity $I_E$ is proportional to the time-averaged energy density $u_E = \epsilon_0\, I_E/2$ (in $\mathrm{J/m}^3$) stored by the fluctuating electric field at a point, where $\epsilon_0$ is the electric permittivity of vacuum [14].

Note that, while the convention taken in Eq. (1) for the definition of the polarization matrix is common in works dealing with polarization theory, the alternative convention $\mathbf{\Phi}_E = \langle \boldsymbol{\varepsilon}^{*}(t)\boldsymbol{\varepsilon}^{\mathrm{T}}(t) \rangle$ (superscript * indicating complex conjugate) is used frequently under the scope of optical coherence theory [12]. In that case the signs of the imaginary parts of the off-diagonal elements of $\mathbf{\Phi}_E$ must be inverted.

The electric degree of polarimetric purity (or degree of polarization) of $\mathbf{\Phi}_E$ is defined as [15-18]

$$P_E = \sqrt{\frac{1}{2}\left(3\,\mathrm{tr}\,\hat{\mathbf{\Phi}}_E^2 - 1\right)} \,, \tag{2}$$

whose value is bounded by $0 \le P_E \le 1$. The lower and upper limits correspond, respectively, to an unpolarized electric state and a pure (fully polarized) state.

The intrinsic Stokes parameters associated with $\mathbf{\Phi}_E$ are given by the 6-tuple $(I_E, I_E P_{El}, I_E P_{Ed}, n_{EO1}, n_{EO2}, n_{EO3})$ [19], where the symbols $P_{El}, P_{Ed}, n_{EO1}, n_{EO2}, n_{EO3}$ represent, respectively, the degree of linear polarization, the degree of directionality (i.e., the stability of the plane of the polarization ellipse), and the three components of the intrinsic representation of the electric spin vector. These six parameters, together with the three orientation angles $(\varphi_E, \phi_E, \theta_E)$ of the intensity ellipsoid [3,19,20], constitute the nine independent real parameters required for a full description of the state encoded in $\mathbf{\Phi}_E$.

We next consider the magnetic polarization matrix, which characterizes the second-order polarization properties of the magnetic field. Let $h_x(t), h_y(t), h_z(t)$ be the analytic signals of the three components of the magnetic field $\mathbf{H}(t)$ in the same reference frame *XYZ* used for the electric field. To represent the elements of the magnetic polarization matrix (defined below) in units $\mathrm{V}^2/\mathrm{m}^2$ consistent with those of the electric matrix, we invoke $\beta_i(t) = Z_0\, h_i(t) \quad (i = x, y, z)$, where $Z_0 = \sqrt{\mu_0/\epsilon_0}$ is the impedance of vacuum, with $\mu_0$ being the magnetic susceptibility of vacuum. The magnetic analytic signal vector is therefore substituted by $\boldsymbol{\beta}(t) = [\beta_x(t), \beta_y(t), \beta_z(t)]^{\mathrm{T}}$, and the magnetic polarization matrix is defined as

$$\mathbf{\Phi}_H = \langle \boldsymbol{\beta}(t) \otimes \boldsymbol{\beta}^{\dagger}(t) \rangle \,. \tag{3}$$

The magnetic intensity is $I_H = \mathrm{tr}\,\mathbf{\Phi}_H$ and we introduce the *magnetic polarization density matrix* $\hat{\mathbf{\Phi}}_H$ via $\mathbf{\Phi}_H = I_H \hat{\mathbf{\Phi}}_H$. The corresponding magnetic degree of polarimetric purity $P_H$ is obtained from Eq. (2) by substituting $\hat{\mathbf{\Phi}}_E$ with $\hat{\mathbf{\Phi}}_H$. Analogously to the electric case, $\mathbf{\Phi}_H$ admits an interpretation in terms of the corresponding set of six magnetic intrinsic Stokes parameters $(I_H, I_H P_{Hl}, I_H P_{Hd}, n_{HO1}, n_{HO2}, n_{HO3})$ and the three orientation angles $(\varphi_H, \phi_H, \theta_H)$ of the magnetic intensity ellipsoid.

The pair $(\mathbf{\Phi}_E, \mathbf{\Phi}_H)$ provides a complete statistical description of the second-order polarization properties of each field component separately, forming the two diagonal blocks of the electromagnetic polarization matrix to be introduced next.

## 3. Electromagnetic polarization matrix

We now introduce the electromagnetic analytic signal vector

$$\mathbf{\Psi}(t) = \begin{pmatrix} \boldsymbol{\varepsilon}(t) \\ \boldsymbol{\beta}(t) \end{pmatrix}, \tag{4}$$

so that the corresponding (Hermitian positive semidefinite) *electromagnetic polarization matrix* is defined as the 6×6 matrix

$$\mathbf{\Phi} = \left\langle \mathbf{\Psi}(t) \otimes \mathbf{\Psi}^{\dagger}(t) \right\rangle = \begin{pmatrix} \mathbf{\Phi}_E & \mathbf{\Phi}_{EH} \\ \mathbf{\Phi}_{EH}^{\dagger} & \mathbf{\Phi}_H \end{pmatrix}. \tag{5}$$

Its 3×3 diagonal blocks, $\mathbf{\Phi}_E$ and $\mathbf{\Phi}_H$, correspond to the electric and magnetic polarization matrices, respectively. The 3×3 off-diagonal blocks, $\mathbf{\Phi}_{EH}$ and $\mathbf{\Phi}_{EH}^{\dagger}$, account for all electric-magnetic cross-correlation properties. The total intensity is given by $I = \mathrm{tr}\,\mathbf{\Phi} = I_E + I_H$.

Note that the choice $\beta_i(t) = Z_0\, h_i(t)$, renders the blocks directly comparable and allows for defining dimensionless electric-magnetic cross-correlation descriptors. In this way, the matrix $\mathbf{\Phi}$ provides a unified representation of electric and magnetic second-order statistics, valid for arbitrary random stationary vector fields in vacuum, including broadband fields. In particular (*i*) the electric and magnetic blocks can be compared directly, (*ii*) the cross-correlation blocks quantify electric–magnetic structure without unit imbalance, (*iii*) all entries correspond to physical observables accessible through field-correlation measurements, and (*iv*) scalar physical descriptors extracted from the matrix, such as degrees of purity and correlation indices, remain dimensionless and invariant under unit conventions.

In certain constrained situations, the 6×6 representation becomes reducible. For a single monochromatic plane wave with a well-defined propagation direction **k**, the magnetic analytic signal is deterministically related to the electric one via Maxwell's equations (e.g., $\mathbf{H} \propto \mathbf{k} \times \mathbf{E}$) [14]. In that case, the magnetic polarization matrix and the cross-correlation blocks are fully determined by the electric polarization matrix, so $\mathbf{\Phi}$ contains redundant information. This reduction does not hold for general superpositions involving multiple propagation directions, near-field/evanescent components, or partial coherence, for which $\mathbf{\Phi}$ generally contains genuinely independent electric, magnetic, and cross-correlation information.

## 4. Spin vectors and physical interpretation

The electric and magnetic spin vectors derived from $\mathbf{\Phi}_E$ and $\mathbf{\Phi}_H$, respectively, are given by [3,21,22]

$$\begin{aligned} \mathbf{n}_E &= 2\,\mathrm{Im}\left(-\Phi_{23},\Phi_{13},-\Phi_{12}\right)^{\mathrm{T}}, \\ \mathbf{n}_H &= 2\,\mathrm{Im}\left(-\Phi_{56},\Phi_{46},-\Phi_{45}\right)^{\mathrm{T}}, \end{aligned} \tag{6}$$

where $\Phi_{kl}$ $(k,l=1,\ldots,6)$ are the elements of $\mathbf{\Phi}$ and Im stands for the imaginary part. Thus, both $\mathbf{n}_E$ and $\mathbf{n}_H$ are axial vectors, i.e., pseudovectors associated with local rotations that do not change sign under spatial inversion, and they are derived from the antisymmetric parts of the electric and magnetic polarization matrices, respectively. The direction and sign of these vectors quantify the chirality (handedness under mirror reflection) of each field **E** and **H**, considered separately.

These polarimetric spin vectors have dimensions of intensity and represent the average magnitude and direction of the axis of rotation of the respective electric and magnetic vectors. Because of the frequency dependence of the monochromatic spectral realizations, the directions of $\mathbf{n}_E$ and $\mathbf{n}_H$ do not, in general, coincide with those of the corresponding spin angular momentum vectors [22].

## 5. Electric-magnetic correlation matrix

The 36 independent parameters that, in the most general case, characterize $\mathbf{\Phi}$, can be interpreted in terms of the nine parameters of $\mathbf{\Phi}_E$, the nine of $\mathbf{\Phi}_H$ (both sets already discussed), and eighteen physical parameters encoded in the cross-correlation matrix $\mathbf{\Phi}_{EH}$.

To interpret $\mathbf{\Phi}_{EH}$, it is useful to analyze its entries in terms of the geometry of the field components they couple. In a Cartesian frame, each element $(\mathbf{\Phi}_{EH})_{ij} = \langle \varepsilon_i(t)\,\beta_j^*(t)\rangle$ with $i,j\in\{x,y,z\}$ quantifies the statistical correlation between the electric component $\varepsilon_i$ and the (rescaled) magnetic component $\beta_j$ (in this section, from now on, time dependence is omitted for brevity). We refer to parallel correlations as those involving components along the same axis, $\langle \varepsilon_i\,\beta_i^*\rangle$ (diagonal elements of $\mathbf{\Phi}_{EH}$), and to orthogonal correlations as those that couple components along mutually perpendicular axes, $\langle \varepsilon_i\,\beta_j^*\rangle$ with $i\neq j$ (off-diagonal elements). In this way, parallel correlations describe how electric and magnetic fluctuations are coupled along a common direction, whereas orthogonal correlations encode the coupling between transverse electric and magnetic components, which underlies quantities such as the local Poynting vector and related directional or chiral properties (studied below).

Each complex correlation $\langle \varepsilon_i\,\beta_j^*\rangle$ can further be characterized by its relative phase. Writing

$$\langle \varepsilon_i\,\beta_j^*\rangle = \left|\varepsilon_i\,\beta_j^*\right| e^{i\,\delta_{ij}}, \tag{7}$$

we say that the correlation is *in phase* when $\delta_{ij}=0$ (mod $2\pi$), meaning that the temporal fluctuations of $\varepsilon_i$ and $\beta_j$ tend to grow and decay synchronously, and *in phase quadrature* when $\delta_{ij}=\pm\pi/2$, i.e., when phases of their fluctuations are shifted by 90º. In practice, in-phase correlations are dominated by their real part, $\mathrm{Re}\langle \varepsilon_i\,\beta_j^*\rangle$, whereas phase-quadrature correlations are dominated by their imaginary part, $\mathrm{Im}\langle \varepsilon_i\,\beta_j^*\rangle$. This joint classification (parallel vs orthogonal, and in-phase vs phase-quadrature) provides a transparent physical picture of the different electric–magnetic coupling mechanisms encoded in $\mathbf{\Phi}_{EH}$. It can be formulated by separating the real and imaginary parts of $\mathbf{\Phi}_{EH}$ as $\mathbf{C}_R = \mathrm{Re}\,\mathbf{\Phi}_{EH}$ and $\mathbf{C}_I = \mathrm{Im}\,\mathbf{\Phi}_{EH}$, respectively, and invoking the following four matrices

$$\begin{aligned} &\mathbf{C}_{RS} = \left(\mathbf{C}_R + \mathbf{C}_{\mathrm{R}}^{\mathrm{T}}\right)/2,\;\; \mathbf{C}_{RA} = \left(\mathbf{C}_R - \mathbf{C}_{\mathrm{R}}^{\mathrm{T}}\right)/2, \\ &\mathbf{C}_{IS} = \left(\mathbf{C}_I + \mathbf{C}_{\mathrm{I}}^{\mathrm{T}}\right)/2,\;\; \mathbf{C}_{IA} = \left(\mathbf{C}_I - \mathbf{C}_{\mathrm{I}}^{\mathrm{T}}\right)/2, \end{aligned} \tag{8}$$

so that

$$\mathbf{\Phi}_{EH} = \mathbf{C}_{RS} + \mathbf{C}_{RA} + i\,\mathbf{C}_{IS} + i\,\mathbf{C}_{IA}. \tag{9}$$

The symmetric matrices $\mathbf{C}_{RS}$ and $\mathbf{C}_{IS}$ satisfy $\mathbf{C}_{RS} = \mathbf{C}_{RS}^{\mathrm{T}}$, $\mathbf{C}_{IS} = \mathbf{C}_{IS}^{\mathrm{T}}$, while the antisymmetric matrices obey $\mathbf{C}_{RA} = -\mathbf{C}_{RA}^{\mathrm{T}}$, $\mathbf{C}_{IA} = -\mathbf{C}_{IA}^{\mathrm{T}}$. Each of these matrices encodes a distinct and measurable physical quantity, as detailed below.

### *5.1 Active flux matrix* $\mathbf{C}_{RA}$

Its three independent parameters coincide (up to a factor of 2) with the components of the real part, $\mathbf{P}_R = \mathrm{Re}\langle \boldsymbol{\varepsilon}\times\boldsymbol{\beta}^*\rangle$, of the averaged complex Poynting vector [23,24]

$$\begin{aligned} \mathbf{P} &= \langle \boldsymbol{\varepsilon}\times\boldsymbol{\beta}^*\rangle \\ &= \left(\left\langle \varepsilon_y\beta_z^*\right\rangle - \left\langle \varepsilon_z\beta_y^*\right\rangle, \left\langle \varepsilon_z\beta_x^*\right\rangle - \left\langle \varepsilon_x\beta_z^*\right\rangle, \left\langle \varepsilon_x\beta_y^*\right\rangle - \left\langle \varepsilon_y\beta_x^*\right\rangle\right)^{\mathrm{T}} \end{aligned} \tag{10}$$

where $\times$ represents the vector product. $\mathbf{P}_R$ governs the directional power flux density of the electromagnetic field. The vector **P** (hence its real and imaginary parts) has dimensions of intensity and its SI Poynting vector version in $\mathrm{W/m^2}$ is recovered as $\mathbf{P}/Z_0$.

In the standard complex Poynting formalism, the real part of **P** represents the time-averaged energy flux, whereas its imaginary part is associated with reactive power and oscillatory exchange of stored electric and magnetic energy, as discussed in detail by Seshadri for dipolar and Gaussian light fields [25,26].

### *5.2 Reactive flux matrix* $\mathbf{C}_{IA}$

Its three independent parameters coincide (up to a factor of 2) with components of the imaginary part $\mathbf{P}_I = \mathrm{Im}\langle \boldsymbol{\varepsilon}\times\boldsymbol{\beta}^*\rangle$ of **P**, which measures the local cyclic exchange of energy (reactive flux) between the electric and magnetic fields. This exchange is directional but does not result in a net flux, instead describing oscillatory storage and release of energy between the electric and magnetic components [25,26]. A nonzero reactive flux can exist even in highly polychromatic or partially coherent fields, provided a finite quadrature correlation exists between orthogonal components of $\boldsymbol{\varepsilon}$ and $\boldsymbol{\beta}$.

### *5.3 Cross-correlation phase alignment matrix* $\mathbf{C}_{RS}$

Its six elements $(\mathbf{C}_{RS})_{ij} = (1/2)[\,\mathrm{Re}\langle \varepsilon_i\beta_j^*\rangle + \mathrm{Re}\langle \varepsilon_j\beta_i^*\rangle]$ provide complete information on in-phase statistical correlations between each component of $\boldsymbol{\varepsilon}(t)$ and each component of $\boldsymbol{\beta}(t)$. This symmetric matrix can always be diagonalized through an orthogonal similarity transformation as $\mathbf{Q}_{RS}^{\mathrm{T}}\mathbf{C}_{RS}\mathbf{Q}_{RS} = (1/2)\,\mathrm{Re}[\,\mathrm{diag}\,(\langle \varepsilon_{x'}\beta_{x'}^*\rangle, \langle \varepsilon_{y'}\beta_{y'}^*\rangle, \langle \varepsilon_{z'}\beta_{z'}^*\rangle)]$, where diag indicates a diagonal matrix. Along the transformed reference axes $X'Y'Z'$ only three paired in-phase correlations survive, while the remaining information is encoded in the three orientation angles determining the orthogonal matrix $\mathbf{Q}_{RS}$. The matrix $\mathbf{C}_{RS}$ therefore quantifies the average alignment between the parallel components of the electric and magnetic fields.

Unlike the active and reactive flux matrices (which are built from orthogonal electric–magnetic correlations and are directly tied to the complex Poynting vector), the matrices $\mathbf{C}_{RS}$ and $\mathbf{C}_{IS}$ collect the symmetric parts of the correlations between parallel components of both fields. They vanish for a single transverse wave, but they can be nonzero in structured or multi-directional fields where some components of $\mathbf{E}$ and $\mathbf{H}$ fluctuate synchronously ( $\mathbf{C}_{RS}$ ) or in phase quadrature ( $\mathbf{C}_{IS}$ ) along the same directions.

*5.4 Cross-correlation quadrature alignment matrix* $\mathbf{C}_{IS}$

A nonzero $\mathbf{C}_{IS}$ represents a distinct channel of electric–magnetic coupling associated with quadrature-phase alignment of parallel components. This channel is complementary to the reactive-flux contribution (which involves orthogonal components), and it can therefore reveal reactive or magnetoelectric structure even when the net reactive flux vanishes by symmetry.

The six elements $(\mathbf{C}_{IS})_{ij} = (1/2)[\,\mathrm{Im}\langle \varepsilon_i \beta_j^* \rangle + \mathrm{Im}\langle \varepsilon_j \beta_i^* \rangle]$ describe the phase-quadrature statistical correlations between the components of $\boldsymbol{\varepsilon}(t)$ and those of $\boldsymbol{\beta}(t)$. This symmetric matrix can always be diagonalized through an orthogonal similarity transformation, which can be written as $\mathbf{Q}_{IS}^{\mathrm{T}} \mathbf{C}_{IS} \mathbf{Q}_{IS} = (1/2)\,\mathrm{Im}[\,\mathrm{diag}\,(\langle \varepsilon_{x''} \beta_{x''}^* \rangle, \langle \varepsilon_{y''} \beta_{y''}^* \rangle, \langle \varepsilon_{z''} \beta_{z''}^* \rangle)]$, so that along the transformed reference axes $X''Y''Z''$ (in general different from $X'Y'Z'$) only three paired quadrature correlations survive (intrinsic in-phase cross-correlations), while the remaining information is then encoded in the three rotation angles determining $\mathbf{Q}_{IS}$. Therefore, $\mathbf{C}_{IS}$ expresses the coherent coupling of parallel electric and magnetic components that are phase-shifted by 90°, a mechanism closely related to reactive energy storage.

*5.5 Chirality content*

The antisymmetric parts of the polarization-matrix blocks carry all information about chirality (whose definition has been recalled in Section 4). In particular, the antisymmetric parts of $\boldsymbol{\Phi}_E$ and $\boldsymbol{\Phi}_H$ yield, respectively, the electric and magnetic spin vectors, $\mathbf{n}_E$ and $\mathbf{n}_H$, whereas the antisymmetric part of $\boldsymbol{\Phi}_{EH}$ encodes the full complex Poynting vector $\mathbf{P}$ (both active and reactive components, $\mathbf{P}_R$ and $\mathbf{P}_I$, respectively), thus providing complete information about chiral electromagnetic coupling.

*5.6 Geometric view: eight-object representation of the electromagnetic polarization matrix*

The electromagnetic polarization matrix $\boldsymbol{\Phi}$ is a 6×6 Hermitian coherence matrix and therefore contains 36 independent real parameters. While its block form already separates electric, magnetic, and cross-correlation contributions (Secs. 2–5), it is interesting to reorganize the same information into a set of geometric objects that can be visualized and interpreted directly.

Following the geometric representation of 3D polarization states (where a 3×3 coherency matrix can be associated with an intensity ellipsoid together with an axial spin vector) [21,27] we can encode the full electromagnetic second-order state at a point by four geometric objects, each combining an ellipsoid and a vector (Fig. 1): the *electric polarization object* (EPO) composed of the electric intensity ellipsoid and the electric spin vector $\mathbf{n}_E$; the *magnetic polarization object* (MPO) composed of the magnetic intensity ellipsoid and the magnetic spin vector $\mathbf{n}_H$; *the active electromagnetic correlation object* (ACO) composed of the in-phase correlation ellipsoid and the active Poynting-flux vector $\mathbf{P}_R$; and *the reactive electromagnetic correlation object* (RCO) composed of the quadrature-phase correlation ellipsoid and the reactive Poynting-flux vector $\mathbf{P}_I$.

More specifically, the real symmetric parts, $\mathbf{S}_E = \mathrm{Re}\,\boldsymbol{\Phi}_E$ and $\mathbf{S}_H = \mathrm{Re}\,\boldsymbol{\Phi}_H$, of the electric and magnetic polarization matrices $\boldsymbol{\Phi}_E$ and $\boldsymbol{\Phi}_H$ define the corresponding electric and magnetic intensity ellipsoids through their semiaxes (arranged in decreasing order) composing the respective diagonal matrices: $\mathrm{diag}\,(a_{Ex}, a_{Ey}, a_{Ez}) = \mathbf{Q}_E^{\mathrm{T}}\, \mathbf{S}_E\, \mathbf{Q}_E$ and $\mathrm{diag}\,(a_{Hx}, a_{Hy}, a_{Hz}) = \mathbf{Q}_H^{\mathrm{T}}\, \mathbf{S}_H\, \mathbf{Q}_H$. Here the proper orthogonal matrices, $\mathbf{Q}_E$ and $\mathbf{Q}_H$ (in general different), determine the respective three orientation angles of each ellipsoid with respect to the Cartesian laboratory reference axes.

The imaginary antisymmetric parts, $\mathrm{Im}\,\boldsymbol{\Phi}_E$ and $\mathrm{Im}\,\boldsymbol{\Phi}_H$, are dual to the electric and magnetic spin vectors $\mathbf{n}_E$ and $\mathbf{n}_H$ (Sec. 4).

As for the geometric representation of the electric-magnetic off-diagonal block $\boldsymbol{\Phi}_{EH}$, the in-phase and quadrature correlation ellipsoids exhibit respective semiaxes determined by the diagonal elements (arranged in decreasing order) of $\mathrm{diag}\,(a_{Rx}, a_{Ry}, a_{Rz}) = \mathbf{Q}_R^{\mathrm{T}}\, \mathbf{C}_{RS}\, \mathbf{Q}_R$ and $\mathrm{diag}\,(a_{Ix}, a_{Iy}, a_{Iz}) = \mathbf{Q}_I^{\mathrm{T}}\, \mathbf{C}_{IS}\, \mathbf{Q}_I$. The proper orthogonal matrices, $\mathbf{Q}_R$ and $\mathbf{Q}_I$ (in general different), determine the respective three orientation angles of each ellipsoid with respect to the Cartesian laboratory reference axes.

Finally, the antisymmetric matrices $\mathbf{C}_{RA}$ and $\mathbf{C}_{IA}$ are dual to the two axial vectors $\mathbf{P}_R$ and $\mathbf{P}_I$, respectively. These vectors represent, in a compact geometric form, the directional balance between radiative transport (active flux) and cyclic exchange of stored energy (reactive flux) at the observation point (Secs. 5.1–5.2).

Thus, the 36 independent parameters involved in $\boldsymbol{\Phi}$ can be interpreted in the following manner: each of the four ellipsoids (real symmetric 3×3 tensors) contributes 6 independent parameters (three semiaxes plus three orientation angles), and each of the four axial vectors contributes 3 parameters. This representation thus provides a complete geometric re-encoding of the electromagnetic polarization matrix, particularly suited for visualization and for identifying rotational invariants (the semiaxes of the four ellipsoids and four vector magnitudes).

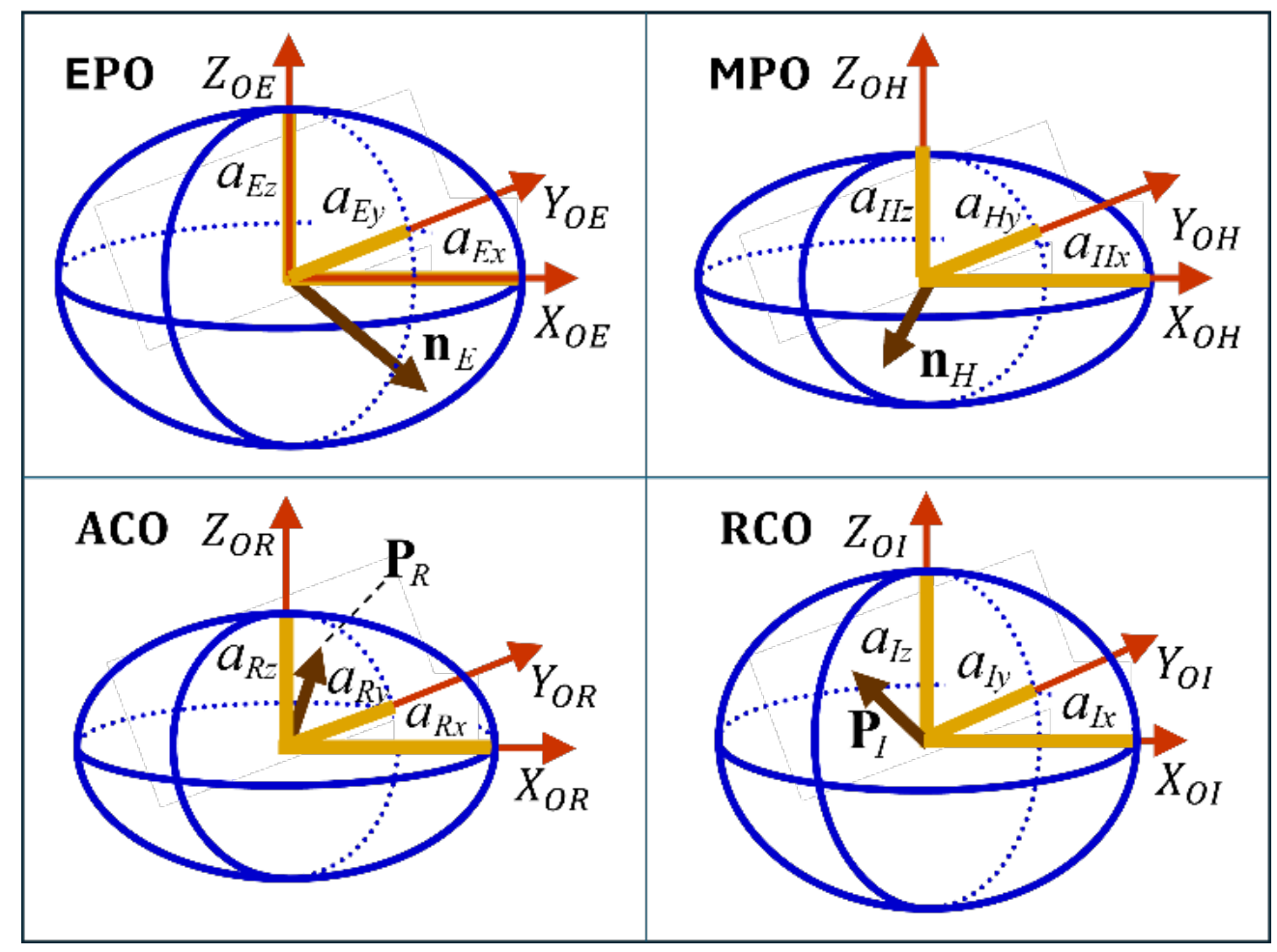

Figure 1. The physical information involved in the electromagnetic polarization matrix $\boldsymbol{\Phi}$ can be interpreted through four geometric objects, each composed of an ellipsoid and an axial vector. All these objects are represented here with respect to their own reference axes: $X_{OE}Y_{OE}Z_{OE}$ of the electric polarization object (EPO), with semiaxes $(a_{Ex}, a_{Ey}, a_{Ez})$; $X_{OH}Y_{OH}Z_{OH}$ of the magnetic polarization object (MPO), with semiaxes $(a_{Hx}, a_{Hy}, a_{Hz})$; $X_{OR}Y_{OR}Z_{OR}$ of the active electromagnetic correlation object (ACO), with semiaxes $(a_{Rx}, a_{Ry}, a_{Rz})$; and $X_{OI}Y_{OI}Z_{OI}$ of the active electromagnetic correlation object (ACO), with semiaxes $(a_{Ix}, a_{Iy}, a_{Iz})$. The four vectors $\mathbf{n}_E$ (electric spin), $\mathbf{n}_H$ (magnetic spin), $\mathbf{P}_R$ (active flux), and $\mathbf{P}_I$ (reactive flux) have specific orientations with respect to the intrinsic axes of the companion ellipsoids.

## 6 Global correlations indices

A set of four scalar indices introduced below summarize, in a compact way, the information contained in the respective four correlation matrices derived from $\boldsymbol{\Phi}_{EH}$: the active flux matrix, the reactive flux matrix, the in-phase alignment matrix, and the quadrature alignment matrix. Each matrix is associated with a specific physical mechanism of electric–magnetic coupling, and its Frobenius norm measures the overall strength of that mechanism. Through appropriate normalization, we obtain dimensionless indices bounded between 0 and 1, which quantify how the available electric–magnetic correlation is partitioned among power transport, reactive exchange of energy, and (in-phase or quadrature) alignment between parallel field components. These indices are defined as follows.

*Active flux index*: $p = \|\mathbf{C}_{RA}\| / \sqrt{I_E I_H} = |\mathbf{P}_R| / \sqrt{2 I_E I_H}$, with $0 \le p \le 1$. The minimum corresponds to zero power flux and the maximum to perfectly in-phase orthogonal electric and magnetic components. Consequently, $p = 1$ implies zero reactive flux $(\mathbf{P}_I = \mathbf{0})$ and vice versa.

The active flux index measures the relative weight of directional power transport encoded in the active flux matrix. Since this matrix is directly associated with the real part of the complex Poynting vector, a vanishing active flux index corresponds to a situation in which there is no net energy flow, even though electric–magnetic correlations may still be present. Conversely, its maximal value is attained when the electric and magnetic fields are correlated essentially as in a pure radiative configuration: orthogonal components of $\boldsymbol{\varepsilon}$ and $\boldsymbol{\beta}$ are perfectly in phase and carry all the available electric–magnetic correlation in the form of a directed power flux. In this extreme case, the other correlation indices vanish, reflecting the fact that no electric–magnetic correlation is left to support reactive exchange or parallel-field coupling.

*Reactive flux index*: $q = \|\mathbf{C}_{IA}\| / \sqrt{I_E I_H} = |\mathbf{P}_I| / \sqrt{2 I_E I_H}$, with $0 \le q \le 1$. The limits correspond, respectively, to zero reactive flux (as for instance a propagating plane wave with $p = 1$) and pure quadrature coupling.

The reactive flux index quantifies the fraction of the total electric–magnetic correlation that is stored in cyclic, non-radiative exchange of energy between the electric and magnetic fields. It is built from the Frobenius norm of the reactive flux matrix, which is, in turn, linked to the imaginary part of the complex Poynting vector. A value of zero indicates that no reactive flux is present, while the upper bound corresponds to a purely reactive situation where all electric–magnetic correlations manifest as quadrature-phase coupling of orthogonal components, without any net energy transport. In such a limit, the active flux index vanishes, emphasizing the complementarity between radiative and reactive mechanisms.

*In-phase alignment index*: $p_{\parallel} = \|\mathbf{C}_{RS}\| / \sqrt{I_E I_H}$, with $0 \le p_{\parallel} \le 1$, measuring the strength of the phase alignment of the mutually parallel electric and magnetic components of the field.

The in-phase alignment index characterizes the strength of in-phase correlations between parallel components of $\boldsymbol{\varepsilon}$ and $\boldsymbol{\beta}$. It is constructed from the Frobenius norm of the symmetric matrix that collects the real parts of the parallel correlations $\langle \varepsilon_{i'} \beta_{i'}^{*} \rangle$. A large value of this index means that, in a suitable orthogonal frame, most of the electric–magnetic coupling can be described as pairs of components along the same axes that fluctuate synchronously in time. This reflects a tendency of $\boldsymbol{\varepsilon}$ and $\boldsymbol{\beta}$ to be statistically aligned in direction and phase, even if the field does not transport energy significantly or exhibits little reactive behavior. In contrast, a small in-phase alignment index indicates that parallel components contribute little to the overall electric–magnetic correlation, which could then be dominated by orthogonal or quadrature-phase couplings.

*Quadrature alignment index*: $p_{\perp} = \|\mathbf{C}_{IS}\| / \sqrt{I_E I_H}$, with $0 \le p_{\perp} \le 1$, representing the magnitude of quadrature-phase alignment between parallel components.

The quadrature alignment index, $p_{\perp}$, plays an analogous role to $p_{\parallel}$, but for parallel components that are in phase quadrature. It is based on the Frobenius norm of the symmetric matrix that collects the imaginary parts of the parallel correlations $\langle \varepsilon_{i'} \beta_{i'}^{*} \rangle$. When this index is large, one can find an orthogonal basis in which $\boldsymbol{\varepsilon}$ and $\boldsymbol{\beta}$ exhibit strong coherent coupling between matching components that are phase-shifted by 90º.

*Quadratic constraint and degree of electric–magnetic correlation.* Because the four matrices derived from $\boldsymbol{\Phi}_{EH}$ form an orthogonal decomposition under the Frobenius inner product, the squares of their norms add up to the squared norm of the full electric–magnetic correlation matrix. This yields a Pythagorean-type relation between the four normalized indices and the overall degree of electric–magnetic correlation. Specifically, the fundamental quadratic constraint

$$\|\boldsymbol{\Phi}_{EH}\|^2 = \|\mathbf{C}_{RA}\|^2 + \|\mathbf{C}_{IA}\|^2 + \|\mathbf{C}_{RS}\|^2 + \|\mathbf{C}_{IS}\|^2, \tag{11}$$

Defines the electric–magnetic correlation. Upon normalization, it leads to

$$c^2 = p^2 + q^2 + p_{\parallel}^2 + p_{\perp}^{\,2}, \tag{12}$$

where $c = \|\boldsymbol{\Phi}_{EH}\| / \sqrt{I_E I_H}$, with $0 \le c \le 1$, is the *degree of electric-magnetic correlation*. Note that, since the signs of phase-related quantities are carried by other observables, e.g., helicity or reactive flux, *c* is taken nonnegative without loss of generality.

Equation (12) shows that the four correlation types represent complementary facets of the same underlying electric–magnetic structure: increasing one of them necessarily reduces the room available for the others at fixed *c*. For a given degree of electric–magnetic correlation, the quadruple $(p, q, p_{\parallel}, p_{\perp})$ lies on a four-dimensional hypersphere of radius *c*, embedded in a five-dimensional conical domain defined by $0 \le c \le 1$. In this geometric picture, pure limiting cases correspond to the tips of the coordinate axes: a configuration with purely radiative coupling has $p = c$ and all other indices zero; a purely reactive configuration has $q = c$; and analogous extreme points exist for purely in-phase or purely quadrature-phase alignment between parallel components.

This construction provides a global, invariant characterization of the electric–magnetic coupling encoded in the electromagnetic polarization matrix: the scalar *c* measures how strongly $\boldsymbol{\varepsilon}$ and $\boldsymbol{\beta}$ are correlated overall, while the four indices specify how that correlation is partitioned among active transport, reactive exchange, and parallel-field alignment in phase or in phase quadrature.

## 7. Polarimetric purity

Let us now consider the electromagnetic polarization density matrix $\hat{\mathbf{\Phi}}$, which can be expressed as follows in terms of four 3×3 submatrices

$$\hat{\mathbf{\Phi}} = \frac{1}{I_E + I_H}\begin{pmatrix} I_E\hat{\mathbf{\Phi}}_E & \sqrt{I_E I_H}\,\hat{\mathbf{\Phi}}_{EH} \\ \sqrt{I_E I_H}\,\mathbf{\Phi}^{\dagger}_{EH} & I_H\hat{\mathbf{\Phi}}_H \end{pmatrix}, \tag{13}$$

where $I_E + I_B = I = \mathrm{tr}\,\mathbf{\Phi}$ and $\hat{\mathbf{\Phi}}_{EH} = \mathbf{\Phi}_{EH}\big/\sqrt{I_E I_H}$. This six-dimensional Hermitian matrix, though not a quantum density matrix in the strict sense, shares the same mathematical properties: positivity, unit trace, and a bounded purity. Observe that the off-diagonal block $\hat{\mathbf{\Phi}}_{EH}$ is not Hermitian and therefore does not possess the formal structure of a density matrix.

### *7.1 Definition of electromagnetic purity*

From the general expression for the degree of purity of $n$-dimensional density matrices [28,29],

$$P_n = \sqrt{\frac{n\,\mathrm{tr}\,\hat{\mathbf{\Phi}}^2 - 1}{n-1}}, \tag{14}$$

the degree of purity of the electromagnetic polarization matrix is defined as

$$P = \sqrt{\frac{1}{5}\left(6\,\mathrm{tr}\,\hat{\mathbf{\Phi}}^2 - 1\right)}. \tag{15}$$

The quantity $\mu = \mathrm{tr}\,\hat{\mathbf{\Phi}}^2$ represents formally the usual purity measure employed in quantum mechanics, bounded as $1/6 \le \mu \le 1$. The lower and upper limits correspond, respectively, to a completely random state (unpolarized and uncorrelated electric and magnetic fields, $P = 0$) and to a pure electromagnetic state (fully polarized electric and magnetic fields with complete cross-correlation, $P = 1$).

### *7.2 Relation between electric, magnetic and total purities*

From Eq. (13), the purity of $\mathbf{\Phi}$ (not to be confused with the degree of polarimetric purity $P$) can be expressed in terms of its submatrices as

$$\begin{gathered} \mu = \frac{1}{I^2}\left[I_E^2\mu_E + I_H^2\mu_H + 2\left\|\mathbf{\Phi}_{EH}\right\|^2\right], \\ \left(\mu_E = \left\|\hat{\mathbf{\Phi}}_E\right\|^2, \mu_H = \left\|\hat{\mathbf{\Phi}}_H\right\|^2\right). \end{gathered} \tag{16}$$

By normalizing the cross-correlation term through the use of the degree of electric-magnetic correlation one obtains

$$\mu = \frac{I_E^2\mu_E + I_H^2\mu_H + 2I_E I_H c^2}{I^2}. \tag{17}$$

This expression shows that the overall purity is a weighted quadratic combination of the electric and magnetic purities, plus the contribution of their mutual correlation. When $I_E = I_H$, Eq. (17) simplifies to

$$\mu = \frac{\mu_E + \mu_H + 2c^2}{4}. \tag{18}$$

In terms of the degrees of polarimetric purity, Eq. (17) takes the form

$$P^2 = \frac{\left(I_E - I_H\right)^2}{5I^2} + \frac{4\left(I_E^2 P_E^2 + I_H^2 P_H^2\right)}{5I^2} + \frac{12 I_E I_H c^2}{5I^2}, \tag{19}$$

where the first term represents an offset of intensity imbalance that disappears when $I_E = I_H$; the second term accounts for the separate contributions of the electric and magnetic purities, and the third term is the contribution of the degree of electric-magnetic correlation. Equation (19) determines the purity-correlation cone of the state (which reduces to an ellipsoid for each feasible fixed value of $P$)

$$\begin{gathered} 1 = \frac{P_E^2}{a^2} + \frac{P_H^2}{b^2} + \frac{c^2}{q^2}, \\ \left[\begin{gathered} a^2 = \frac{r^2}{4I_E^2},\ b^2 = \frac{r^2}{4I_H^2},\ q^2 = \frac{r^2}{12 I_E I_H}, \\ r^2 \equiv 5P^2 I^2 - \left(I_E - I_H\right)^2. \end{gathered}\right] \end{gathered} \tag{20}$$

Note that only the positive octant of the $(P_E, P_H, c)$ space is physically realizable, since the degrees of purity and the degree of electric-magnetic correlation are defined as non-negative normalized quantities. When $I_E = I_H$ Eq. (19) adopts the simplified form

$$P^2 = \frac{P_E^2 + P_H^2 + 3c^2}{5}. \tag{21}$$

Total purity implies that the electric and magnetic polarization matrices are both pure and the fields are perfectly cross-correlated $(1 = P_E = P_H = c)$. Conversely, fully unpolarized electromagnetic states $(\mu = 1/6,\ P = 0)$, for which $\mathbf{\Phi}$ is proportional to the 6×6 unit matrix, entail unpolarized electric and magnetic fields $(P_E = P_H = 0)$ and consequently zero electric-magnetic cross-correlation $(c = 0)$.

These relations show that the electromagnetic purity encompasses both internal (electric or magnetic) purity and mutual coherence, providing a rigorous bridge between polarization and general coherence properties. Thus, the electromagnetic polarization matrix provides a natural framework for identifying mixed states—those in which the electric and magnetic fields are individually partially or totally polarized but only partially correlated—as well as fully coherent pure electromagnetic states. This concept plays an analogous role to entanglement and separability in quantum theory, though in a classical context.

### *7.3 Upper bound of μ fixed by the electric and magnetic purities*

From the Cauchy–Schwarz inequalities associated with positive semidefinite matrices, $\left|\mathrm{tr}(\mathbf{AB})\right|^2 \le \mathrm{tr}\,\mathbf{A}^2\,\mathrm{tr}\,\mathbf{B}^2$, demonstrated by Horn and Mathias [30], it follows that the upper bound for the degree of electric-magnetic correlation is determined by

$$c^2 \le \sqrt{\mu_E \mu_H}. \tag{22}$$

This inequality establishes the physically intuitive result that the strength of the electric–magnetic coupling cannot exceed the product of the intrinsic purities of the individual fields. In terms of the electric and magnetic degrees of polarimetric purity Eq. (22) becomes

$$c^2 \le \frac{1}{3}\sqrt{\left(1 + 2P_E^2\right)\left(1 + 2P_H^2\right)}. \tag{23}$$

When the equality holds, $\mathbf{\Phi}_{EH}$ reaches its maximum physically admissible correlation, meaning that the electric and magnetic fluctuations are as strongly coupled as allowed by their individual purities. In this limit the Cauchy–Schwarz bound is saturated, although the maximum value $c_{\max}$ of $c$ may still be lower than unity when the subfields are mixed. From Eq. (17), we obtain

$$\mu_{\max} = \frac{\left(I_{\mathrm{E}}\sqrt{\mu_{\mathrm{E}}} + I_{\mathrm{H}}\sqrt{\mu_{\mathrm{H}}}\right)^2}{I^2}. \tag{24}$$

For the special case of equal intensities and maximal cross-correlation ($I_E = I_H$, $c^2 = \mu_E = \mu_H = 1$) Eq. (17) yields the limiting condition $\mu = 1$, i.e., a pure electromagnetic state.

*7.4 Purity-correlation space*

According to Eqs. (20) and (23), the space of physically admissible electromagnetic polarization states is completely determined by the following five parameters: the three independent purity parameters $P$, $P_E$, $P_H$, the degree of electric–magnetic correlation $c$, and the relative intensity imbalance parameter $t \equiv |I_E - I_H|$.

A convenient and physically transparent way to explore the four–dimensional space $P$, $P_E$, $P_H$, $c$ is to construct the states in a sequential manner: ($a$) fix arbitrary values for the purities $P_E$, $P_H$, with $0 \le P_E, P_H \le 1$; ($b$) determine the maximum physically allowed (positive) value of the correlation index $c_{\max} = \sqrt[4]{(1+2P_E^2)(1+2P_H^2)}\big/\sqrt{3}$ [Eq. (23)]; ($c$) select an admissible value of $c \le c_{\max}$ and then compute the total electromagnetic purity $P$ from Eq. (20). The result is unique once the parameter $t$ is specified. When $t \neq 0$, this leads to an offset, so that the minimum possible $P$ is $P_{\min} = t\big/\sqrt{5}\,I$. When $t = 0$, $P$ lacks such a lower limit.

An equivalent, and more geometrically intuitive, representation of the physically accessible states is obtained by viewing Eq. (20) as defining a family of ellipsoids in the $(P_E, P_H, c)$ space. For fixed intensity imbalance $t$ and fixed global degree of purity $P$, Eq. (20) represents the positive octant of an ellipsoid centered at the origin whose semiaxes are given by $a$, $b$ and $q$. The inequality (23) acts as a physical boundary condition, limiting the allowed values of $c$ for each pair $(P_E, P_H)$.

The surface $c = c_{\max}(P_E, P_H)$ thus plays the role of a curved "cap" that truncates the family of algebraic ellipsoids. Only the portion of each ellipsoid lying below this surface corresponds to physically realizable electromagnetic states; the region above violates the positive semidefiniteness of $\mathbf{\Phi}$.

Consequently, the full set of realizable states forms a truncated volume in the positive octant determined by $0 \le P_E, P_H, c$ and bounded by: the lower surface $c = 0$ (the constant-$P$ section being an ellipse); the upper surface $c = c_{\max}(P_E, P_H)$ [maximally correlated states compatible with the fixed values of $(t, P_E, P_H)$]; the outermost ellipsoid corresponding to the maximal attainable total degree of purity $P$.

This geometrical picture provides an immediate insight: increasing $P_E$ or $P_H$ expands the ellipsoid along the horizontal axes; increasing $c$ moves the state upward along the correlation axis; but the Cauchy–Schwarz surface ultimately limits the vertical extension of the ellipsoid, thereby fixing the maximum achievable correlation and, consequently, the maximum possible $P$ for given purities $P_E$ and $P_H$.

**8. Electromagnetic polarization properties of two crossing beams with common linear electric polarization**

To illustrate the interest and necessity of the complete electromagnetic polarization formalism introduced above, we consider a simple and physically realizable configuration in free space: two plane waves propagating along mutually orthogonal directions ($Z$ and $Y$, respectively), each with arbitrary statistical properties (random stationary light in general) and arbitrary mutual correlation. These beams are arranged such that their electric fields are linearly polarized along a common axis (the $X$ direction), as shown in Fig. 2. For simplicity, both beams are assumed to have equal intensity $(I = I_1 = I_2)$.

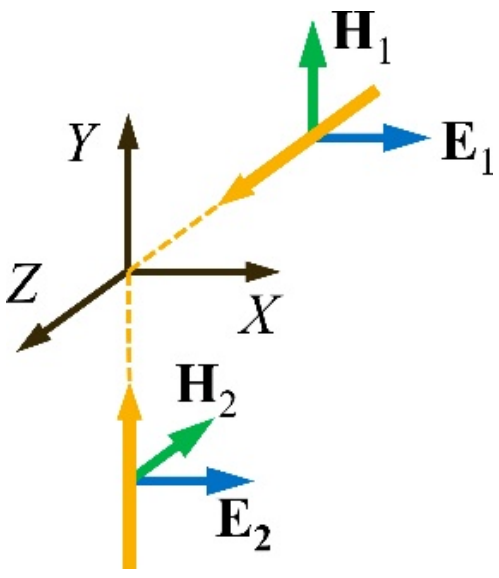

Figure 2. Configuration of the electric and magnetic fields of two plane waves, each with arbitrary spectral profile and arbitrary mutual correlation, propagating along orthogonal directions ($Z$ and $Y$ axes), with linearly polarized electric fields fluctuating along the $X$ axis. The resulting state at the crossing point exhibits different electric and magnetic polarization states.

*8.1 Field representation*

Since for each plane wave the magnitude of $\boldsymbol{\beta}$ is equal to the electric field, the electromagnetic analytic signal vectors of the individual beams are

$$\begin{aligned} \mathbf{\Psi}_1(t) &= \left(\varepsilon_1(t), 0, 0, 0, \varepsilon_1(t), 0\right)^{\mathrm{T}}, \\ \mathbf{\Psi}_2(t) &= \left(\varepsilon_2(t), 0, 0, 0, 0, -\varepsilon_2(t)\right)^{\mathrm{T}}, \end{aligned} \tag{25}$$

so that the analytic signal vector of the composite field is

$$\mathbf{\Psi}(t) = \left(\varepsilon_1(t) + \varepsilon_2(t), 0, 0, 0, \varepsilon_1(t), -\varepsilon_2(t)\right)^{\mathrm{T}}. \tag{26}$$

The electromagnetic polarization matrix then takes the form (zero entries represented by dots for compactness, and time dependence implicitly assumed)

$$\mathbf{\Phi} = \begin{pmatrix} \left\langle \begin{matrix} \varepsilon_1\varepsilon_1^* + \varepsilon_2\varepsilon_2^* \\ +\varepsilon_1\varepsilon_2^* + \varepsilon_2\varepsilon_1^* \end{matrix} \right\rangle & \ldots & \left\langle \varepsilon_1\varepsilon_1^* + \varepsilon_2\varepsilon_1^* \right\rangle & -\left\langle \varepsilon_2\varepsilon_2^* + \varepsilon_1\varepsilon_2^* \right\rangle \\ \vdots & \vdots & \vdots & \vdots \\ \left\langle \varepsilon_1\varepsilon_1^* + \varepsilon_1\varepsilon_2^* \right\rangle & \ldots & \left\langle \varepsilon_1\varepsilon_1^* \right\rangle & -\left\langle \varepsilon_1\varepsilon_2^* \right\rangle \\ -\left\langle \varepsilon_2\varepsilon_2^* + \varepsilon_2\varepsilon_1^* \right\rangle & \ldots & -\left\langle \varepsilon_2\varepsilon_1^* \right\rangle & \left\langle \varepsilon_2\varepsilon_2^* \right\rangle \end{pmatrix}. \tag{27}$$

By applying $\left\langle \varepsilon_1\varepsilon_1^* \right\rangle = \left\langle \varepsilon_2\varepsilon_2^* \right\rangle = I/2$ (equal intensities) and introducing the (complex) degree of mutual correlation between both electric fields $\gamma = 2\left\langle \varepsilon_1\varepsilon_2^* \right\rangle\big/I$ ($|\gamma| \le 1$), then

$$\mathbf{\Phi} = \frac{I}{2}\begin{pmatrix} 2(1+\mathrm{Re}\,\gamma) & \ldots & 1+\gamma^* & -(1+\gamma) \\ \vdots & \vdots & \vdots & \vdots \\ 1+\gamma & \ldots & 1 & -\gamma \\ -(1+\gamma^*) & \ldots & -\gamma^* & 1 \end{pmatrix}. \tag{28}$$

Note that in this configuration the resulting electric and magnetic intensities are, in general, different:

$I'_E = I\,(1+\mathrm{Re}\,\gamma)$, $I'_H = I$. The equality $I'_E = I'_H$ holds only when $\mathrm{Re}\,\gamma = 0$. This illustrates that, for superpositions leading to a non-plane-wave (as in this example), electric and magnetic second-order statistics are not necessarily identical, and the cross-correlation block $\mathbf{\Phi}_{EH}$ is essential to describe the energy-flow and other cross-correlation properties.

*8.2 Polarimetric analysis*

The structure of $\mathbf{\Phi}$ evidences that, even for such a simple configuration, the electric and magnetic submatrices differ markedly. The electric polarization matrix corresponds to a linearly polarized (pure) electric state, whereas the magnetic matrix exhibits a variety of forms depending on $\gamma$. Specifically, for $\gamma = 0$, the magnetic matrix represents a 2D unpolarized magnetic state, since the magnetic contributions of both beams are statistically independent and mutually perpendicular; for $\gamma = 1$ (perfect correlation), the magnetic field becomes linearly polarized along the bisector of the *Y* and *Z* axes, with an orientation at 45° in the *YZ* plane; for $\gamma = \pm i$ (purely imaginary), the magnetic field components are in quadrature phase, resulting in a circularly polarized magnetic state; and for a generic complex $\gamma = |\gamma| e^{i\phi}$, the magnetic field exhibits elliptical polarization whose ellipticity and handedness are governed by the amplitude and phase of $\gamma$.

Thus, by adjusting the correlation amplitude and phase between the two beams, the magnetic polarization state can be continuously tuned from linear to circular. Consequently, the electric spin vector vanishes in this configuration $(\mathbf{n}_E = \mathbf{0})$ since the electric field is strictly linear. However, the magnetic spin is generally nonzero and given by $\mathbf{n}_H = I\,(\mathrm{Im}\,\gamma, 0, 0)^{\mathrm{T}}$. This property enables controlled generation or modulation of the magnetic spin by varying the mutual phase between the electric fields of the two beams.

The average complex Poynting vector of the composite field is $\mathbf{P} = I\,(0, 1+\gamma, 1+\gamma^*)^{\mathrm{T}}/2$. Hence, the net power flow (active flux) lies along the bisector of the *Y* and *Z* axes in the YZ plane, while the direction of the reactive flux is orthogonal to it within the *YZ* plane and its magnitude is controlled by $\mathrm{Im}\,\gamma$.

As for the purity–correlation space determined by Eqs. (20) and (23), this two-beam configuration explores only the vertical plane defined by a perfectly pure electric state, $P_E = 1$. For each fixed modulus $|\gamma|$ of the mutual coherence between the beams, varying the phase $\phi$ of $\gamma$ traces a one-parameter curve lying on an ellipsoid of constant total electromagnetic purity. Along this curve there is a specific value of the phase for which the electric and magnetic fields become uncorrelated $(c = 0)$, touching the lower admissible boundary of the plane, and another phase for which the electric–magnetic correlation saturates the Cauchy–Schwarz inequality $c^2 \leq P_E P_H$ (reaching the upper "Cauchy–Schwarz cap"). As $|\gamma|$ increases from 0 to 1, the magnetic purity $P_H$ rises monotonically from 1/2 (2D unpolarized magnetic state) to 1 (pure magnetic state), and the total electromagnetic purity approaches unity, indicating that the composite field tends to a fully coherent, deterministically polarized configuration.

*8.3 Experimental and conceptual implications*

This simple configuration demonstrates that, although Maxwell's equations couple **E** and **H** at the field level, the corresponding constraints act primarily on two-point correlation functions (Wolf equations) [12,13]. Therefore, the local one-point second-order polarization matrices $\mathbf{\Phi}_E$ and $\mathbf{\Phi}_H$ are not, in general, algebraically determined by one another, except under additional assumptions (e.g., single-direction plane-wave fields or paraxiality).

The example also reveals that the electric–magnetic cross-correlation matrix $\mathbf{\Phi}_{EH}$ plays a crucial role in describing states where both fields are partially correlated. In fact, $\mathbf{\Phi}_{EH}$ is the block that determines how the electric and magnetic polarization matrices are tied together: it fixes the overall degree of electric–magnetic correlation, sets the complex Poynting vector (and thus whether the available correlation manifests as active power flow or reactive exchange of energy), and controls the chiral coupling between $\boldsymbol{\varepsilon}$ and $\boldsymbol{\beta}$. As a result, different electromagnetic states that share the same $\mathbf{\Phi}_E$ and $\mathbf{\Phi}_H$ can only be distinguished at the level of $\mathbf{\Phi}_{EH}$. This highlights the distinction from deterministic six-component representations: while a single six-component field vector can be used to construct quadratic observables for a given realization, only the full 6×6 polarization matrix captures partial polarization and partitions the electric–magnetic coupling into the four correlation channels discussed above.

A fully analogous experiment can be envisioned where the magnetic fields of two orthogonally propagating waves are linearly polarized along a common direction, interchanging the roles of the electric and magnetic components. In that case, the electric spin becomes controllable via the imaginary part of the corresponding correlation coefficient.

Such configurations could be implemented experimentally using intersecting laser beams or microwave fields with controlled coherence, enabling direct observation of magnetic spin modulation or reactive flux phenomena. They also exemplify how the electromagnetic polarization matrix provides a unified quantitative framework for describing and engineering coupled electric–magnetic polarization states.

## 9. Concluding remarks

In summary, the electromagnetic polarization matrix introduced here provides a compact and self-consistent second-order framework that unifies the electric and magnetic polarization matrices with all their cross-correlations. Beyond the conventional electric-field description, this 6×6 formalism delivers a complete statistical representation of polarization states in random stationary and polychromatic fields, and it does so in a way that is directly interpretable in terms of measurable quantities.

A key message is that, even in simple configurations, electric and magnetic second-order properties need not coincide for superpositions that depart from a single plane wave. In such cases, the cross-correlation block becomes essential: it encodes the full structure of electric–magnetic coupling, clarifies when energy transport and reactive exchange occur, and identifies the cross-correlation properties associated with in-phase and quadrature alignments. The worked example of two orthogonally propagating beams with a common linear electric polarization illustrates these points in a physically realizable setting and shows how magnetic polarization and spin can be tuned by the phase and degree of correlation between the electric fields.

From a structural viewpoint, the global degree of electromagnetic purity is governed by a purity–correlation relation that combines (*i*) the electric and magnetic degrees of purity, $(P_E, P_H)$, (*ii*) the degree of electric–magnetic correlation $c$, and (*iii*) the intensity imbalance parameter $t$. Constant-purity sections form ellipsoids in the $(P_E, P_H, c)$ space, truncated by the Cauchy–Schwarz cap that sets the physically admissible upper bound for $c$. Only the positive octant is accessible, since the three indices are defined as non-negative normalized quantities. This geometry offers a clear, quantitative picture of how the different contributions cooperate to produce the observable electromagnetic order and of the ultimate limits imposed by physical consistency.

Conceptually, the 6×6 matrix plays the role of a classical density matrix: it is Hermitian, positive semidefinite, and of unit trace after normalization. This analogy supports a rigorous discussion of separable-like versus non-separable (maximally correlated) electromagnetic states in terms of the rank and factorability of the cross-correlation block (while remaining firmly within classical statistical optics). Operationally, the formalism suggests concrete experimental routes (e.g., intersecting beams or structured fields) to control electric–magnetic coupling, modulate reactive flux, and engineer spin/helicity in random or polychromatic conditions.

Looking ahead, although in this work we have restricted attention to random fields in free space, the same second-order framework can be reformulated in the frequency domain and thereby extended to propagation in material media, where the electromagnetic polarization matrix is defined at each frequency using the appropriate constitutive relations. In this spectral setting, the approach naturally covers near-field and evanescent regimes, and it lends itself to computational estimation of the cross-correlation block from measurable field data. It also opens an avenue to compare electromagnetic non-separability with quantum-inspired criteria at the level of second-order statistics, potentially enriching both polarization theory and coherence theory with a unified language.

**Funding.** Research Council of Finland (349396, 354918, PREIN 346518). B. Fazlpour acknowledges financial support from the Finnish Ministry of Education and Culture through the Quantum Doctoral Education Pilot Program (QDOC VN/3137/2024-OKM-4).

**Conflicts of interest.** The authors have nothing to disclose.

**Data availability**. No data were generated in this study.